AUTHOR ACCEPTED MANUSCRIPT

# The Tragedy of the Cognitive Commons: How AI Could Disrupt the Regeneration of Professional Expertise

Nolan Lovett

NATO Special Operations University, Belgium

ORCID: https://orcid.org/0009-0003-1982-830X

Author note. The author was concurrently a doctoral candidate at Old Dominion University, Norfolk, Virginia, USA, during the research and writing of this article.

Correspondence. nlove002@odu.edu

# The Tragedy of the Cognitive Commons: How AI Could Disrupt the Regeneration of Professional Expertise

## Abstract

Artificial intelligence is reshaping cognitive work, but Human Resource Development scholarship has treated this transformation as an organizational training challenge, leaving the collective regeneration of professional expertise unexamined. This conceptual paper introduces the Cognitive Commons framework, integrating commons theory, HRD scholarship, and distributed cognition to explain how rational AI adoption decisions can deplete the shared expertise pool professions require for renewal. The framework distinguishes Internalized Mastery (deep domain knowledge from sustained practice) from Distributed Mastery (orchestrating human-AI systems), and develops the Validation Tether: effective AI oversight depends on the expertise AI adoption may undermine. Early labor market and clinical evidence suggests possible disruption to expertise-regeneration pathways in highly AI-exposed sectors, though adoption is recent and the strongest signals come from leading sectors rather than all professions. Five factors determine occupational vulnerability, and governance arrangements may form across organizational, professional-association, and policy levels. The paper reframes expertise development as collective stewardship rather than organizational optimization, with implications for HRD theory and workforce policy.

## Introduction

In 1968, Garrett Hardin described how rational individuals, acting independently in their own self-interest, inevitably deplete shared resources upon which all depend. His articulation of the Tragedy of the Commons has shaped thinking about environmental conservation, public goods, and collective action problems for over half a century (Hardin, 1968). Today, as artificial intelligence (AI) systems rapidly assume cognitive tasks across professional domains, we face a parallel tragedy, one that Human Resource Development (HRD) has not yet recognized as a collective action problem, despite accumulating evidence of its emergence.

This paper invokes Hardin's metaphor for its structure, not its fatalism. Hardin (1968) presented commons depletion as the near-inevitable result of rational individual action, but later scholarship, above all Ostrom's (1990, 2010) empirical work on enduring common-pool institutions, showed that communities often sustain shared resources through governance arrangements such as boundary definition, monitoring, graduated sanctions, and collective choice. The Cognitive Commons is therefore not presented as doomed. The tragedy framing identifies a collective action problem that current arrangements leave unaddressed; the governance question, taken up later, is whether institutions can be designed to sustain the regeneration pathways on which professional expertise depends.

The argument that follows rests on a distinction in evidential status. The dissociation between assisted performance and independent capability is empirically established, directly measured wherever studies test unaided performance after AI exposure (Budzyń et al., 2025; Macnamara et al., 2024). The profession-level depletion this paper describes is a structural prediction from that dissociation and commons theory, not an observed outcome, and it is offered as a falsifiable account of where current incentives lead rather than a claim that the tragedy has already arrived.

The HRD field is grappling with AI's implications for workplace learning and expertise development. Scholarship in this space addresses how AI transforms training delivery and skill requirements (Ardichvili et al., 2024; Chai et al., 2025; Ekuma, 2024; Khandelwal et al., 2024) and what ethical frameworks should govern organizational AI adoption (Chang & Ke, 2024; Chen, 2024; Yorks & Jester, 2024). This emerging body of work raises concerns about responsible AI implementation, human agency, and the need for HRD to guide organizational adoption.

Much of this scholarship frames AI's challenge in terms of workforce reskilling and upskilling, emphasizing the need for workers to develop new competencies such as AI literacy, prompt engineering, and human-AI collaboration skills. Khandelwal et al. (2024) describe how "modern AI-based HRD systems leverage machine learning algorithms to tailor training programmes to employees' unique needs and preferences" (p. 630), emphasizing personalized learning for AI-augmented work. Ekuma (2024) notes that AI adoption has "precipitated a paradigmatic shift in the skills requirements for HRD professionals," requiring "expertise in data analytics, technology management, and digital literacy" (p. 215). This framing positions HRD as facilitating worker adaptation to AI through competency development interventions.

Organizations do need workers who can collaborate effectively with AI systems, and reskilling initiatives deserve continued investment. Alone, however, they are insufficient responses to the deeper structural problem AI adoption can create. Wiles et al.'s (2024) experimental evidence is instructive: when AI assistance was provided during a skill-building task, participants performed significantly better during the access period, but this performance advantage did not transfer to subsequent unassisted performance. AI-assisted productivity and AI-independent expertise are not the same capability, and developing one does not automatically build the other. Human factors researchers have recognized this paradox for decades: Bainbridge (1983) demonstrated that automation simultaneously increases the need for human skill and reduces the opportunity to develop it, an insight extended to system-level analysis by Hollnagel and Woods (2005). What this literature has not addressed is the collective action structure that emerges when every organization in a profession faces this paradox simultaneously. In the most AI-exposed sectors, organizations are already eliminating the entry-level positions that have historically provided the foundation for expertise development (Brynjolfsson et al., 2025; Hampole et al., 2025), and these disruptions continue to be interpreted as temporary workforce adjustments rather than as signals of regeneration failure. Ardichvili (2022) warned that automation creates limited opportunities for developing deep expertise, but the field has not developed frameworks to understand this as a collective resource problem requiring governance at the professional level.

Understanding how AI adoption creates commons dynamics invisible to organizational-focused analysis requires integrating literatures that have remained disconnected: commons theory from environmental studies and public policy, expertise development research from HRD and adult education, distributed cognition theory from cognitive science, and systems thinking. This synthesis follows Jaakkola's (2020) framework for integrating literatures fragmented across disciplines, bridging these domains. Conceptual papers provide foundational construct development distinct from formal theory-building, a process Rocco et al. (2022) describe as involving conceptual development of key constructs and their relationships (p. 127) that precedes operationalization and empirical testing.

This paper introduces the Cognitive Commons as the collective pool of deep human expertise within a profession that functions as a shared resource whose regeneration can no longer be taken for granted. Understanding how this tragedy unfolds requires recognizing that AI-era work demands two forms of expertise: Internalized Mastery, representing deep domain knowledge concentrated within individual minds through sustained deliberate practice, and Distributed Mastery, representing fluency in orchestrating intelligence distributed across human-AI systems. These forms serve complementary functions, but commons depletion jeopardizes both. The Validation Tether explains their interdependence: effective Distributed Mastery fundamentally depends on Internalized Mastery as its foundation, because substantive validation of AI outputs requires the deep domain knowledge that traditional developmental pathways build. This conceptual foundation enables future research to operationalize these constructs, develop testable propositions using established theory-building frameworks (Lynham, 2002), and examine governance mechanisms in specific professional contexts.

This paper proceeds in four movements. First, it defines the Cognitive Commons and establishes when and where commons dynamics operate, integrating evidence from labor economics, clinical medicine, and experimental psychology. Second, it synthesizes commons theory, HRD scholarship, distributed cognition, and systems thinking to derive the Validation Tether: effective Distributed Mastery depends on the Internalized Mastery that AI adoption can undermine. Third, it formalizes these relationships in an integrative model of commons depletion. Finally, it argues that HRD must expand beyond organizational expertise optimization to encompass profession-level commons stewardship.

## The Cognitive Commons Construct

The Cognitive Commons is defined as the collective pool of deep human expertise within a profession, encompassing the distributed reservoir of professionals who possess internalized domain knowledge, tacit understanding, robust mental models, and the judgment capabilities necessary to perform complex cognitive work independently and respond adaptively to novel situations outside algorithmic parameters. These experts perform substantive validation of automated system outputs, recognizing domain-specific errors, identifying contextually inappropriate recommendations, and detecting subtle flaws that superficially coherent outputs may contain. This commons functions as a shared resource upon which all organizations operating in a domain depend, even when they do not directly employ every expert within the pool.

The Cognitive Commons represents a significant extension of how HRD conceptualizes expertise. While Swanson (1995) grounded HRD in the development of expertise at individual, process, and organizational levels, and while scholars have examined expertise development extensively within organizational contexts (Cherrstrom & Bixby, 2018; Grenier & Kehrhahn, 2008), the field has given insufficient attention to expertise as a collective resource that exists at the profession or societal level. The Cognitive Commons framework suggests that HRD must think systemically not just about organizational systems (Yawson, 2013), but about the health of entire expertise ecosystems that transcend individual organizational boundaries.

**Distinguishing the Cognitive Commons from Adjacent Constructs**

The Cognitive Commons differs from several adjacent constructs in ways that motivate the new term. Human capital, in both its individual and organizational forms, locates expertise in persons or firms (Becker, 1964); the Cognitive Commons is profession-level, a shared stock that no single actor owns. Communities of practice describe the social processes through which practitioners learn (Lave & Wenger, 1991); the Cognitive Commons describes the resulting stock, its regeneration, and its depletion, not the learning process itself. Workforce capacity refers to the availability of workers and can be restored through recruitment; the Cognitive

Commons refers to the deep, validation-capable expertise that recruitment can redistribute but cannot regenerate on demand. Professional expertise names a competence held by an individual (Dreyfus et al., 1986; Ericsson, 2006); the Cognitive Commons names the collective pool of such competence and the developmental system that renews it. The construct earns its place by directing attention to a profession-level stock-and-regeneration dynamic that none of these adjacent terms makes visible.

**Why This Constitutes a Commons: Three Defining Characteristics**

Commons theory is particularly apt for understanding expertise threatened by the dynamics emerging in AI-exposed sectors because it captures three structural characteristics that create vulnerability to collective action failure. These characteristics explain why individually rational organizational decisions can produce collectively problematic outcomes despite no organization intending such results.

First, the Cognitive Commons is collectively dependent. Organizations in any domain rely on hiring experts from a shared talent pool, creating a free-rider problem: they benefit from available expertise while each has incentive to avoid bearing the cost of developing it, free-riding on others' developmental investments. This challenge is heightened in interconnected fields like healthcare and finance, where the existence of expertise in the broader ecosystem, even when not directly employed, is crucial. For instance, a financial institution's AI risk assessment relies on human experts for validation, patient safety in hospitals depends on clinicians maintaining diagnostic skills, and the reliability of AI-generated code hinges on programmers being able to audit and identify bugs. In these domains, collective expertise provides regulatory capacity, crisis response capability, and knowledge spillovers that benefit all organizations regardless of whether they directly employ those experts. This creates what economists term a public goods problem (Samuelson, 1954), where the benefits of collective expertise availability are non-excludable (all organizations benefit from profession-wide expert capacity) while the costs of developing new experts are borne by specific organizations that invest in prolonged developmental pathways.

Second, the commons is non-exclusive. Organizations cannot easily prevent others from benefiting from its existence. The general availability of deep expertise in a profession provides collective insurance against systemic failures, regulatory oversight capability, and knowledge transfer channels that benefit all organizations regardless of their individual contributions to expertise development. A firm that eliminates entry-level positions and relies entirely on AI can still benefit from the profession's aggregate expert capacity when hiring experienced practitioners trained by other organizations, consulting external specialists whose expertise developed through traditional pathways, or depending on regulatory oversight by domain experts whose validation capability the profession maintains. This non-excludability creates free-rider incentives (Olson, 1965), where rational actors seek to benefit from commons resources (the profession-wide expert pool and its validation capacity) without contributing to commons maintenance through investments in developmental infrastructure.

Third, the commons is degradable through overexploitation of the regeneration mechanism. Organizations can extract value from the commons (by deploying AI that depends on expert validation while eliminating the entry-level positions that train validators) potentially faster than the commons can regenerate. Unlike traditional commons such as fisheries or forests where depletion becomes visible through declining catches or bare hillsides, Cognitive Commons degradation manifests through generational employment patterns rather than immediate resource scarcity. Existing experts continue validating AI outputs and managing exceptions competently, creating an impression of stability while the regeneration mechanism experiences disruption in affected sectors. The erosion becomes measurable through cohort-level labor market analysis showing systematic elimination of early-career positions (Brynjolfsson et al., 2025; Hampole et al., 2025), but organizations and policymakers commonly interpret these patterns as temporary labor market adjustment (workers displaced by automation finding alternative opportunities) rather than recognizing the potential commons regeneration failure they may represent. This gap between visible symptoms and recognized underlying dynamics compounds the governance challenge, as the collective action structure driving potential depletion remains largely unacknowledged even as empirical indicators emerge in leading sectors.

The emerging pattern in highly AI-exposed sectors thus represents the commons analogue of Hardin's (1968) overgrazing problem. Each organization that eliminates an entry-level position captures 100% of the efficiency gains (reduced salary costs, eliminated training expenses, faster task completion through AI) while distributing the expertise depletion cost across all organizations that depend on profession-wide expert availability. The rational calculation favors appropriation over contribution. Organizations that invest in developmental infrastructure, whether preserving traditional pipelines or building new ones, face competitive disadvantage against those that rely on hiring already-trained experts from the shrinking pool. Market forces reward efficiency optimization and penalize organizations that bear disproportionate costs for expertise development whose benefits accrue profession-wide.

As Ostrom (1990) demonstrated through extensive empirical research on natural resource commons, such collective action problems need not inevitably lead to tragedy if governance mechanisms enable coordinated stewardship. The Cognitive Commons currently lacks adequate governance mechanisms in the most vulnerable sectors, particularly those without strong professional licensing infrastructure. The governance mechanisms Ostrom identified as essential for commons sustainability, including boundary definition, monitoring, graduated sanctions, and collective choice arrangements, remain largely absent in the most vulnerable professional sectors.

**Applicability: When and Where the Commons Dynamic Operates**

The Cognitive Commons framework applies when three structural conditions co-occur. First, AI must perform tasks that have historically served as primary developmental contexts for the expertise in question, specifically the early-career cognitive labor and production tasks through which novices build domain knowledge, tacit understanding, and mental models. Professions where novices develop expertise primarily through embodied practice insulated from AI substitution, such as nursing or the building trades, exhibit lower commons vulnerability. Second, those developmental pathways must be difficult to replicate outside organizational work contexts. Where formal education currently substitutes effectively for workplace experience in building the relevant expertise, commons vulnerability is moderated. Third, the profession must exhibit the three commons characteristics: collective dependence on a shared expertise pool, non-excludability of access to that pool, and vulnerability to regeneration failure as organizations eliminate entry-level positions.

The framework's structural conditions are not specific to any national context. The commons dynamic operates wherever AI substitutes for developmental cognitive labor within professions characterized by permeable labor markets, collective dependence on shared expertise, and organizational rather than educational developmental pathways. The early empirical evidence reviewed in this paper draws on US (Brynjolfsson et al., 2025; Hampole et al., 2025), UK (Klein Teeselink, 2025), and German (Engberg et al., 2025) labor markets, reflecting the current state of the research base rather than a theoretical limitation. Whether the dynamics manifest differently under varying institutional arrangements, such as strong apprenticeship traditions, regulated labor markets, or different professional licensing regimes, represents an empirical question the framework generates rather than a boundary it assumes. The framework also assumes a particular cultural model of professional formation, in which expertise develops through entry-level employment in hierarchical organizations and is governed through formal credentialing institutions. This model dominates Western professional sectors and the evidence base reviewed here, but it does not exhaust the ways human expertise has been or can be transmitted. Apprenticeship traditions, guild structures, and community-based knowledge transmission represent alternative arrangements that may exhibit different commons dynamics or different vulnerability to AI substitution.

The appropriate unit of analysis for commons dynamics is the occupation rather than the national economy. An occupation's commons vulnerability depends on the degree to which AI substitutes for the specific cognitive labor that historically built expertise in that field, not on aggregate labor market conditions. Software engineering exhibits different commons vulnerability than plumbing, and within software engineering, backend infrastructure development may differ from front-end development depending on which tasks AI substitutes most readily. This occupational specificity generates testable predictions: professions should show greater commons disruption where AI exposure indices are higher, where early-career developmental pathways have historically been concentrated in organizational rather than educational contexts, and where the three commons characteristics are most pronounced.

The boundary conditions also clarify what the framework excludes. It does not address professions facing wholesale disintermediation, where AI eliminates the coordination or translation function the profession exists to perform rather than depleting its expertise pipeline; these represent a distinct phenomenon requiring separate analysis. The framework also assumes that AI systems continue to require human oversight for reliable performance in consequential domains; if AI achieves fully autonomous reliability, the collective action problem shifts from expertise depletion to workforce displacement, a qualitatively different challenge. Several countervailing dynamics and five structural factors determine differential vulnerability across professions that meet these conditions.

Countervailing dynamics moderate commons depletion. AI adoption may create new developmental pathways that partially compensate for eliminated positions; collaborative AI workflows requiring active clinical engagement have improved diagnostic accuracy rather than degrading it (Everett et al., 2025). More broadly, the aggregate labor market evidence is mixed, with studies documenting adaptive capacity (Manning & Aguirre, 2026), considerable heterogeneity in retrainability across occupations and labor market conditions (Hyman et al., 2025), and null effects on earnings and recorded hours among workers in Denmark during the first two years of generative AI adoption (Humlum & Vestergaard, 2025). These findings establish that the depletion mechanism is not universal.

The framework identifies five factors that determine an occupation's commons vulnerability. First, task substitutability: the degree to which AI can perform the specific cognitive tasks that have historically provided developmental experiences for novices. Where AI substitutes for high-substitutability tasks, such as routine information synthesis, standard document production, and basic data analysis, developmental pathways face greater disruption than where AI augments rather than substitutes. Second, regulatory intensity: professions with stringent requirements for documented human expertise and supervised practice, such as medicine, law, and engineering, face institutional pressures that partially counteract commons depletion even as market forces favor AI substitution. Third, safety criticality: professions where expertise failures produce immediately visible catastrophic consequences create accountability structures that resist degradation of substantive validation capability. Fourth, professional association strength: occupations with strong professional associations capable of enforcing developmental standards have institutional infrastructure for commons stewardship that others lack. Fifth, work modularization: the degree to which professional work can be unbundled into discrete AI-compatible tasks versus requiring integrated domain judgment across the entire workflow.

These five factors generate testable predictions about differential commons vulnerability. Software engineering, financial analysis, and legal research exhibit high task substitutability, relatively low regulatory intensity compared to medicine, and work modularization that facilitates AI substitution, placing them in higher vulnerability categories. Medicine and engineering exhibit higher regulatory intensity and safety criticality that create institutional counterpressures, suggesting slower but not absent degradation dynamics.

This analysis has direct implications for where HRD scholarship and governance attention should concentrate. Rather than treating the Cognitive Commons as a claim about universal expertise degradation, the framework directs attention toward specific occupations, contexts, and conditions where commons dynamics operate most strongly. Research and governance interventions should target high-vulnerability occupations first, while comparative research across vulnerability levels can test the framework's predictions empirically. The goal is not to resist AI adoption broadly but to govern it in ways that preserve developmental infrastructure in the occupations and contexts where commons dynamics make such preservation most urgent.

The regeneration mechanism that AI adoption disrupts operates through organizational developmental pipelines: entry-level hiring creates opportunities for supervised practice on progressively complex tasks, through which novices develop the tacit knowledge, pattern recognition, and judgment that characterize deep expertise (Dreyfus et al., 1986; Ericsson, 2006). This workplace learning dominates formal training in producing professional expertise (Eraut, 2004; McCall et al., 1988), operating through what Lave and Wenger (1991) term legitimate peripheral participation. Systematic reviews of career and technical education confirm the pattern: such programs improve initial employment but show no measurable impact on long-term earnings (Lindsay et al., 2024), suggesting that workplace developmental contexts, not educational preparation, drive expertise formation. This regeneration system functions effectively across most professional domains today; the question is whether disruption patterns in leading sectors signal dynamics that will spread more widely.

The developmental pathways through which Internalized Mastery forms are not reducible to exposure or time on task. Deep expertise develops through scaffolded engagement with progressively complex problems, in which practitioners struggle with difficulty, monitor and revise their own reasoning, and repeatedly correct flawed assumptions against evidence (Ericsson, 2006; Mezirow, 1991; Rind, 2022). The cognitive struggle is not incidental friction to be engineered away; it is the mechanism through which schemas, diagnostic judgment, and metacognitive control are built. Assistance that removes the struggle, rather than scaffolding the practitioner through it, can leave performance intact while arresting the development of the capability that performance is meant to signal.

**Economic Foundations of Commons Vulnerability**

From an economic perspective, the Cognitive Commons represents a market failure in human capital investment. Becker's (1964) seminal work distinguished between firm-specific human capital (valuable only to the current employer) and general human capital (transferable across firms). Firms have clear incentives to invest in specific capital but systematically under-invest in general capital because trained workers can defect to competitors, preventing the investing firm from capturing full returns (Becker, 1964). This creates what economists term a positive externality in training investment, where social returns exceed private returns because benefits partially accrue to other firms when workers change employers.

The deep expertise constituting Internalized Mastery represents quintessentially general human capital. A financial analyst's domain knowledge, a physician's diagnostic capability, or a software engineer's architectural understanding transfers across firms. Traditional human capital theory predicts chronic under-investment in such general training because firms bear full development costs while benefits distribute across competitors who can hire trained workers. Labor economists have documented that despite this market failure, some training investment occurs because labor market imperfections (wage compression, information asymmetries, monopsony power) can make general training investment privately rational under certain conditions (Acemoglu & Pischke, 1998; Stevens, 1994).

AI adoption transforms this under-investment problem into a regeneration challenge. Classical human capital theory assumes that despite under-investment, some training occurs because firms need trained workers to perform tasks, and the work itself provides developmental opportunities. This created a regeneration system where organizations pursuing private goals (filling positions, completing tasks) produced the collective benefit (profession-wide expertise pool). AI eliminates this productive activity, creating a second-order market failure: not merely under-investment in training, but elimination of the productive activity through which expertise development naturally occurred as a byproduct of employment. The prior equilibrium was not the product of governance; it was the product of accidental alignment. Organizations maintained developmental pipelines not because they recognized any stewardship obligation to the profession, but because they needed entry-level workers to perform entry-level work. The operational necessity of junior labor was the hidden governance mechanism: commons regeneration occurred as a side effect of normal business operations, not as the product of collective will or institutional design. AI makes this latent collective action problem manifest.

**Early Disruption in AI-Exposed Sectors: First Signals of Regeneration Failure**

These commons characteristics create vulnerability to collective action failure that is no longer merely theoretical. In the most AI-exposed occupations, measurable disruption to the regeneration mechanism may be emerging. Widespread generative AI adoption represents an extremely recent development, with most organizational implementation beginning in late 2022. The majority of professions show no measurable disruption to traditional hiring patterns. The evidence presented here captures early-stage dynamics in a specific subset of highly AI-exposed sectors.

Brynjolfsson et al. (2025) analyzed payroll data from a provider covering over 25 million U.S. workers, drawing on an analysis sample of 3.5 to 5 million workers per month, examining employment trends by age and AI exposure following widespread generative AI adoption. Their findings reveal a stark pattern in the most AI-exposed occupations: workers ages 22 to 25 experienced a 16% relative decline in employment between October 2022 and September 2025, even after controlling for firm-level shocks. In contrast, employment for workers aged 35-49 in those same occupations grew by over 8% over the same period. Workers in less AI-exposed occupations showed comparable employment growth across all age groups. The age-specific pattern, in other words, reflects AI adoption in particular sectors rather than broader economic trends. Complementary analysis by Hampole et al. (2025) using 58 million LinkedIn profiles reinforces this cohort-specific pattern, demonstrating that age-specific employment declines reflect systematic elimination of developmental positions rather than temporary labor market adjustment.

These employment patterns reflect two distinct yet reinforcing mechanisms through which AI disrupts commons regeneration. The first mechanism, direct position elimination, operates when AI systems autonomously perform entry-level tasks, allowing organizations to eliminate those positions entirely. The labor market evidence primarily captures this first mechanism; the 16% employment decline for workers ages 22-25 represents positions that simply no longer exist.

The second mechanism, augmentation without internalization, operates even when entry-level positions persist. Junior workers using AI achieve productivity levels that historically required years of experience, enabling organizations to accomplish the same work with fewer entry-level employees. More critically, those who are hired develop fluency in AI orchestration while skipping the cognitive struggle through which Internalized Mastery forms. This mechanism proves harder to detect because employment numbers may appear healthy while regeneration quality silently degrades. The Brynjolfsson et al. (2025) data captures Mechanism 1 effectively but cannot detect Mechanism 2, underscoring why qualitative research examining developmental experiences is essential alongside quantitative labor market analysis.

Brynjolfsson et al. (2025) further demonstrate that employment declines concentrate in occupations where AI primarily automates work rather than augments human judgment, domains where AI systems independently perform tasks that once provided learning opportunities.

Several limitations warrant acknowledgment. The data span less than three years of widespread generative AI adoption, representing early-stage rather than mature dynamics. The patterns concentrate in the most AI-exposed occupations and do not characterize professional employment broadly. The evidence does not demonstrate widespread validation failures or profession-wide expertise collapse.

Additional empirical evidence documents three pathways through which AI adoption undermines independent capability. First, practitioners who rely heavily on AI assistance show measurably reduced independent performance. Budzyń et al. (2025) found that endoscopists who adopted AI-assisted detection systems showed reduced independent accuracy compared to baseline, and Natali et al.'s (2025) cross-specialty review documents consistent patterns of reduced independent diagnostic performance under AI adoption. Rinta-Kahila et al.'s (2023) organizational case study documented similar self-reinforcing skill erosion cycles in an accounting firm. Second, AI-assisted performance does not transfer to unassisted contexts. Wiles et al. (2024) found no significant advantage on unassisted assessment after AI-assisted training, and Kumar et al. (2025) and Bangerl et al. (2025) found limited skill carryover from AI-assisted ideation tasks, with the effect most pronounced on tasks requiring domain judgment. Daley (2025) arrives at structurally identical predictions through mathematical modeling: when AI assistance is expected to persist, rational workers underinvest in human capital. Third, workers who retain validation capacity are already failing to exercise it. Niederhoffer et al. (2025) found that 40% of surveyed full-time employees received substantively flawed AI-generated content in the past month, with recipients spending nearly two hours on average addressing each instance, and Benzing et al. (2025) found that 60% of employees reported feeling confident enough in AI output that they do not routinely check its accuracy. As practitioners abandon disciplined validation habits, the conditions for intergenerational transfer of substantive validation culture erode alongside the formal developmental pathways.

The Cognitive Commons remains healthy and functional across most professional domains today. The question is whether the disruption patterns now visible in leading sectors represent early warning signals of dynamics that, if unaddressed, will spread more widely and ultimately compromise the regeneration mechanisms that have sustained professional expertise for generations.

**Why Rational Organizations Deplete the Commons: The Overgrazing Mechanism**

The overgrazing mechanism described above operates through what economists term a negative externality (Pigou, 1920): each organization captures private efficiency gains from AI adoption while the cost of reduced expertise regeneration distributes across all organizations that depend on collective expert availability.

The commons tragedy becomes visible when examining how organizations plan to meet their ongoing need for deep expertise after eliminating developmental infrastructure. Rational organizations reason: "We don't need expensive entry-level positions and prolonged developmental pathways. When we need deep expertise (for validation, crisis management, or novel situations) we'll hire experienced workers from the labor market." This market-based appropriation strategy works for individual organizations precisely because it free-rides on the commons. The experienced experts available for hire today exist because other organizations made developmental investments 5-20 years ago: hiring entry-level workers, supervising them through extended learning curves, providing progressively complex work, and bearing the risk that trained workers might leave for competitors. The time delay between cause and effect obscures the commons depletion. If organizations eliminate entry-level positions beginning in 2023, the labor market for experienced workers appears healthy because it reflects developmental investments from 2003-2020. The impact of foregone developmental investments will not manifest in experienced worker availability until 2030-2045, long after the organizational decisions that caused the shortage.

The free-rider problem emerges because no individual organization's decision meaningfully affects profession-wide expert availability. When an organization eliminates entry-level positions in a profession with hundreds of thousands of workers, the impact on collective expert capacity appears negligible. However, when many organizations reason identically, aggregate effects may prove severe despite each individual decision appearing rational and inconsequential. This free-rider problem is amplified by what is termed here as the Human Reserve Paradox. Organizations need expertise held in reserve for validation, crisis management, and novel situations that exceed AI capabilities. But they lack sufficient economic incentive to maintain that reserve when costs fall on them individually while benefits distribute across the ecosystem. Traditional economic models cannot adequately price the Cognitive Commons because its value remains latent until crisis reveals its absence. Like insurance, its worth becomes evident only when needed. Unlike insurance, no market mechanism exists to internalize commons maintenance costs across beneficiaries. As all organizations reason identically, the regeneration mechanism erodes while appropriation from the shrinking pool continues unabated.

## Theoretical Synthesis

Understanding the Cognitive Commons requires integrating literatures that have remained disconnected: commons theory, HRD and adult education scholarship, distributed cognition, and systems thinking (Jaakkola, 2020). Each is necessary; none is sufficient. The preceding sections have deployed each literature individually. This section identifies what emerges from their combination.

Commons theory provides the analytical vocabulary demonstrated above: the stock-regeneration distinction, collective action failure, free-rider dynamics, and governance design principles (Hardin, 1968; Ostrom, 1990). Applied to expertise, commons theory reveals why apparent stability masks crisis: the expert stock appears robust while the regeneration mechanism erodes. What commons theory alone cannot explain is why this particular commons proves so vulnerable. Natural resource commons regenerate when harvesting stops. Professional expertise does not, because the developmental requirements are specific and non-substitutable.

That specificity is what HRD and adult education scholarship provides. HRD rests on psychological, economic, and systems foundations (Swanson & Holton, 2009), but the field has primarily applied these within organizational boundaries (Swanson, 1995). As the regeneration analysis established, expertise develops through informal workplace learning (Eraut, 2004; Blume et al., 2010), legitimate peripheral participation (Lave & Wenger, 1991), transformative learning (Mezirow, 1991), and sustained deliberate practice (Ericsson, 2006). The economic foundations demonstrate why AI transforms chronic underinvestment in these pathways into elimination of the developmental mechanism itself. HRD scholarship specifies the content of what is being depleted; commons theory explains the collective action structure driving the depletion. Neither alone captures both dimensions.

Systems thinking provides the analytical architecture that connects the preceding literatures into a predictive framework (Senge, 1990; Yawson, 2013). The stock-regeneration distinction central to the commons analysis is a stock-flow concept from systems dynamics. The accidental alignment argument describes structural coupling: organizational hiring needs and professional expertise regeneration were coupled through the operational necessity of junior labor, and AI decouples them. The six-node causal chain (Figure 1) formalizes these relationships, making visible the reinforcing feedback loops and time delays that prevent individual actors from recognizing their collective impact on the regeneration mechanism.

Distributed cognition research poses the most direct challenge to the framework's premises. If cognitive capability can distribute across human-AI networks (Hutchins, 1995), why does concentrated individual expertise matter? Hutchins's naval navigation analysis showed that effective coordination requires humans who understand what distributed systems are doing and why, not merely how to operate them. Extended through recent research on transactive memory in human-AI partnerships (Bienefeld et al., 2023; Woolley & Gupta, 2024), this reveals what is termed the Validation Tether: effective orchestration of opaque AI systems requires independent domain understanding sufficient to recognize when systems operate outside competence boundaries (Hollnagel & Woods, 2005). Distributed cognition, rather than undermining the argument, identifies the mechanism through which Internalized Mastery remains essential even in AI-augmented work.

The Cognitive Commons emerges as a recognizable phenomenon only when these literatures combine. Commons theory identifies the collective action structure. HRD scholarship specifies the developmental content that is vulnerable. Systems thinking reveals the cross-scale dynamics that make depletion invisible until crisis. Distributed cognition establishes why the expertise being depleted remains indispensable.

The mechanism through which AI attenuates expertise formation is best understood as redistribution, not simple elimination. AI does not only remove cognitive work; it reallocates it, shifting effort away from the generative struggle that builds Internalized Mastery and toward the orchestration, comparison, and evaluation that constitute Distributed Mastery (Clark & Chalmers, 1998; Hutchins, 1995). Whether that reallocation builds or erodes expertise is conditional on how the work is designed. Where AI assistance preserves the practitioner's own attempt and requires them to evaluate, question, and justify, the new cognitive work can itself be developmental; where it supplies finished outputs that bypass that effort, augmentation attenuates Internalized Mastery formation (Rind, 2026). Evidence that the depletion mechanism operates conditionally rather than universally (Everett et al., 2025; Hyman et al., 2025; Manning & Aguirre, 2026) reinforces the point: the question for HRD is not whether AI weakens expertise but under what design conditions it builds or erodes it.

**The Validation Tether: Why Distributed Mastery Depends on Internalized Foundations**

The Validation Tether represents the fundamental dependence of effective AI orchestration on independent domain understanding. This mechanism operates through a critical asymmetry: while AI systems can perform tasks that appear correct, humans require deep domain knowledge to recognize when superficially coherent outputs contain substantive errors.

The tether manifests at two distinct validation levels. Surface validation involves detecting obvious errors, checking logical coherence, and assessing general plausibility. This capability can develop through training and experience with AI systems themselves. Workers learn to spot formatting inconsistencies, recognize nonsensical outputs, and identify responses that violate basic constraints. Surface validation represents important quality control but proves insufficient for professional work where errors often hide beneath plausible surfaces.

Substantive validation requires recognizing domain-specific errors, identifying contextually inappropriate recommendations despite technical correctness, and detecting subtle flaws that superficially coherent outputs may contain. This capability fundamentally depends on the internalized domain knowledge, robust mental models, and tacit understanding that constitute Internalized Mastery. A financial analyst needs deep market knowledge to recognize when algorithmically generated risk assessments contain flawed assumptions despite mathematical correctness. A physician needs clinical expertise to identify when AI diagnostic recommendations ignore patient-specific factors despite matching symptom patterns. A software engineer needs architectural understanding to detect when AI-generated code will create maintenance problems despite executing correctly.

This dependence reflects a long-standing observation about the structure of human knowledge. Hayek (1945) argued that economically consequential knowledge is largely dispersed and tied to particular circumstances of time and place, resisting capture in any centralized representation. Polanyi (1966) made the parallel epistemological claim that human knowing exceeds what can be articulated. The Validation Tether is the application of this insight to human-AI systems. Surface validation operates on what AI can articulate; substantive validation requires the tacit, contextual knowing that articulation cannot capture. AI orchestration thus depends on the very form of expertise that AI cannot itself transmit.

The empirical evidence demonstrates that surface validation proves inadequate when AI produces plausible but flawed outputs. Vicente and Matute (2023) found that participants exposed to biased AI recommendations reproduced those biases in their own subsequent judgments, even after AI assistance was removed. Critically, 80.7% of participants detected errors in the AI recommendations, demonstrating functional surface validation. But they continued following erroneous advice, suggesting that detecting a problem and overriding an algorithmic recommendation require different cognitive capacities. Surface validation functioned; substantive validation failed. The participants knew something was wrong but could not articulate what or why, leaving them unable to override the algorithmic output.

Dell'Acqua et al. (2026) documented the same pattern among elite consultants. Productivity gains materialized on tasks within AI capability frontiers, but performance declined on tasks outside those boundaries, precisely when substantive validation became critical. The consultants could not reliably distinguish between situations where AI assistance would enhance performance and situations where it would degrade it. This discrimination capability requires the deep domain understanding that enables practitioners to assess whether a given problem falls within or outside the AI system's competence envelope.

Substantive validation depends not only on what practitioners know but on how they regard knowing. Epistemic beliefs, the assumptions practitioners hold about the nature, source, and justification of knowledge (Hofer & Pintrich, 1997), govern whether they interrogate an AI output or defer to it. The distinction is visible in the Vicente and Matute (2023) result already noted: most participants detected the flawed recommendation, but they deferred to it anyway, treating the system as an authority rather than as a source of claims to be checked. Practitioners who treat AI outputs as authoritative tend to accept fluent answers without scrutiny, while those with more sophisticated epistemic orientations verify, cross-reference, and question before accepting (Rind et al., 2026). Substantive validation therefore has an epistemic precondition as well as a cognitive one: the domain knowledge that makes an error recognizable, and the epistemic stance that prompts the practitioner to look. This compounds the developmental concern, because the depleted pathways that erode domain knowledge also remove the apprenticeship contexts in which practitioners learn to question authority and test claims against evidence.

The validation tether creates decisive implications for commons dynamics. Organizations pursuing AI adoption eliminate entry-level positions that provide the deliberate practice, progressive complexity, and cognitive struggle through which Internalized Mastery develops. Effective Distributed Mastery requires exactly that internalized foundation to function reliably. Workers who develop fluency in AI orchestration without acquiring deep domain knowledge gain productivity on routine tasks while losing the substantive validation capability that prevents errors on non-routine tasks. They cannot recognize when AI operates outside its competence boundaries because they lack the domain understanding necessary for such recognition.

This interdependence means that commons depletion threatens both forms of expertise simultaneously. By eliminating developmental pathways to Internalized Mastery, organizations undermine the foundation upon which effective Distributed Mastery depends. The tragedy is that organizations need what they are destroying: the deep expertise that enables workers to know when to trust AI outputs, when to modify them, and when to override them entirely. Without substantive validation capability, increased AI adoption paradoxically increases rather than decreases organizational risk, as workers gain confidence in their ability to deploy AI while losing the domain knowledge necessary to recognize its failures.

## An Integrative Model of Commons Depletion

The preceding analysis converges on a six-node causal chain that describes how AI adoption can produce collective expertise depletion despite no organization intending such outcomes. The first three nodes capture the mechanisms through which developmental pathways erode. First, AI adoption eliminates entry-level positions through Mechanism 1: organizations directly remove developmental roles because AI performs those functions autonomously. Second, AI assistance enables junior workers who remain employed to achieve productivity without cognitive struggle, Mechanism 2, through which augmentation without internalization attenuates expertise formation. Third, both pathways reduce Internalized Mastery formation across successive cohorts while appearing, to organizational accounting, as efficiency gains. The next three nodes capture the systemic consequences. Fourth, depleted Internalized Mastery degrades the Validation Tether, as workers who did not develop deep domain knowledge through traditional pathways lack the foundational understanding necessary for substantive validation. Fifth, degraded validation capability increases systemic risk as AI adoption scales, because the human oversight function that should catch AI errors when they occur becomes increasingly unreliable. Sixth, systemic risk remains latent until crisis events reveal it, because surface validation continues to function even as substantive validation erodes, sustaining organizational confidence that AI adoption is performing well. Figure 1 presents this six-node causal chain visually.

**Figure 1**

*The Causal Chain of Cognitive Commons Depletion*

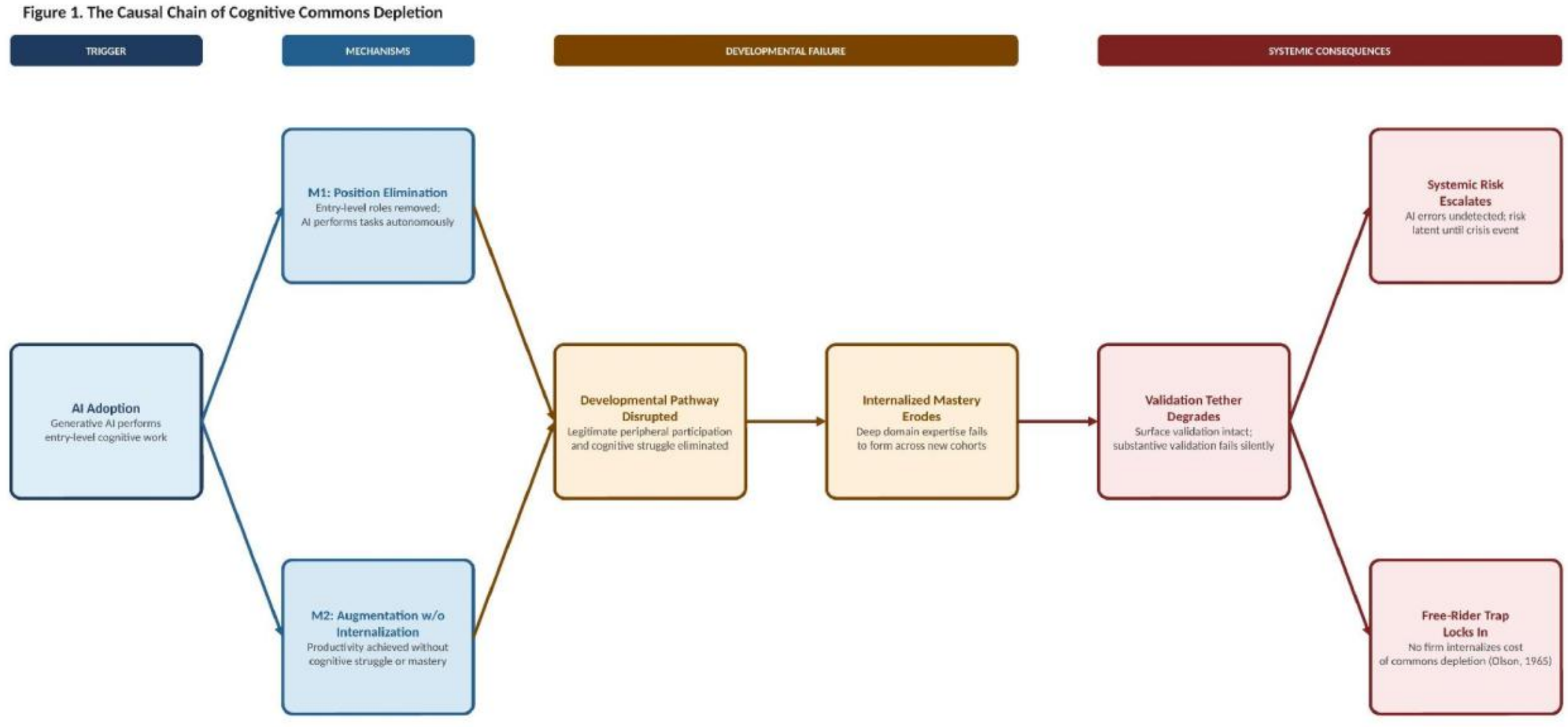

Figure 1. The Causal Chain of Cognitive Commons Depletion

This integrative model clarifies that commons depletion operates through two simultaneous failure modes that compound one another. Stock depletion occurs as the overall count of practitioners who possess deep Internalized Mastery declines, a function of the generational replacement dynamic in which new cohorts enter professions without having traversed the traditional developmental pathways that built expertise in their predecessors. Functionality degradation occurs as practitioners who nominally possess senior titles or years of experience show shallower domain expertise than their counterparts a generation earlier, because even those hired at entry level increasingly developed fluency in AI orchestration rather than in the underlying domain. These failure modes are distinct in timing: stock depletion becomes numerically measurable through cohort analysis over 10 to 20 year periods, while functionality degradation may manifest in validation error rates on shorter timescales. Table 1 presents operational definitions of the six core constructs comprising the framework, together with their key distinguishing attributes.

**Table 1**

*Construct Definitions for the Cognitive Commons Framework*

| Construct | Definition | Key Distinguishing Attribute |
|---|---|---|
| Cognitive Commons | The collective pool of deep professional expertise within a field that functions as a shared resource maintained through ongoing professional practice and degradable through failure of its regeneration mechanism. | Unlike individual human capital (Becker, 1964), non-rivalrous in access but collectively degradable. Differs from communities of practice (Lave & Wenger, 1991) in emphasizing resource-stock properties rather than social learning processes. Not equivalent to workforce capacity, which can be restored through recruitment. |
| Internalized Mastery | Deep domain knowledge, cognitive schemas, diagnostic judgment, and pattern recognition that develops through sustained cognitive struggle and authentic professional practice across progressive levels of complexity. | Requires cognitive struggle during formation; cannot be acquired through observation or AI-assisted performance alone. Distinguishes from Distributed Mastery in providing the independent domain foundation required for substantive validation. Distinguishes from general expertise by its public-goods dimension. |
| Distributed Mastery | Proficiency in orchestrating AI systems to produce professional-quality outputs, including prompt engineering, output curation, and human-AI workflow design within a specific domain. | Productive output achievable without deep domain internalization. Differs from general AI literacy (Khandelwal et al., 2024) in its domain-specific orientation. Does not provide the independent domain competence required for substantive validation. |
| Validation Tether | The dependence of effective AI orchestration on practitioner Internalized Mastery; the mechanism through which AI outputs become substantively evaluable rather than merely surface-plausible. | Distinguishes surface validation from substantive validation as two qualitatively different oversight functions. Its degradation is invisible to performance metrics that track output volume rather than output validity. Differs from human oversight (generic) in making the Internalized Mastery dependence theoretically explicit. |
| Surface Validation | Verification that AI outputs are internally consistent, properly formatted, and plausible in surface features; does not require independent domain competence. | Accessible without Internalized Mastery; compatible with Distributed Mastery alone. Permits substantively incorrect outputs to pass undetected. Differs from |

| | | substantive validation in expertise requirement. |
|---|---|---|
| Substantive Validation | Verification that AI outputs are domain-specifically correct, including detection of outputs that appear plausible but contain errors identifiable only through independent domain expertise. | Requires Internalized Mastery; distinguishes competent human oversight from its performance. Degrades alongside Internalized Mastery across successive practitioner cohorts. Documented empirically in error propagation under AI adoption (Budzyń et al., 2025; Vicente & Matute, 2023). |

***Note.*** Constructs are defined as they function within the Cognitive Commons framework. Internalized Mastery and Distributed Mastery are presented as analytically distinct forms of expertise that may co-occur within individual practitioners. Validation Tether refers to the structural relationship between these forms, not to a trainable disposition.

## Implications for HRD Theory and Practice

The Tragedy of the Cognitive Commons demands fundamental reconceptualization of HRD's scope, methods, and theoretical foundations. This section outlines three critical areas requiring attention: measuring commons health, designing governance mechanisms, and reconceptualizing HRD education.

### Specific Contributions to HRD Theory

The framework makes three contributions to HRD theory. First is a contribution of level of analysis: it moves HRD's account of expertise from the individual and the organization to the profession-level ecosystem in which expertise is collectively held and renewed, extending

National HRD's concern with system-level capability (Garavan et al., 2021; McLean, 2004). Second is a developmental contribution: it specifies how AI adoption may disrupt the work-based pathways through which expertise is regenerated, locating the threat not in any single training decision but in the cumulative erosion of the developmental infrastructure itself. Third is a governance contribution: it positions HRD as a potential steward of that developmental infrastructure rather than only a provider of training interventions, raising the question of how professions might monitor and sustain the health of their expertise commons (Ostrom, 1990). Together these reframe HRD's object from individual capability to the collective system that produces it.

**Measuring and Monitoring Commons Health**

First, we urgently need to develop observable indicators of commons health and depletion. HRD scholarship documents AI's impact on specific populations and processes. Ekuma's (2024) systematic review identified critical gaps, noting the significant lack of longitudinal studies investigating AI's lasting impact on HRD processes and outcomes, and calling for research tracking these effects over time (p. 218). However, current research largely examines organizational-level outcomes (training effectiveness, employee performance, skill development) rather than profession-level expertise ecosystem health. Rather than asking "How effectively do organizations use AI in training?" we must ask "Is the profession maintaining sufficient developmental infrastructure to regenerate deep expertise across successive cohorts?" .

Methodological templates already exist, including Hampole et al.'s (2025) task-level AI exposure measures from resume data, Vicente and Matute's (2023) experimental paradigms distinguishing task completion from substantive validation capability, and Macnamara et al.'s (2024) analysis of cognitive skill erosion.

The commons framework advances claims at three distinct evidential levels that researchers should distinguish when building on this work. Empirically grounded claims rest on published evidence from the studies reviewed: that labor market data show cohort-specific early-career employment declines in AI-exposed occupations (Brynjolfsson et al., 2025; Hampole et al., 2025; Klein Teeselink, 2025); that experimental and clinical evidence documents access-dependent rather than transferable performance gains from AI assistance (Wiles et al., 2024; Budzyń et al., 2025; Kumar et al., 2025; Bangerl et al., 2025); and that behavioral evidence documents widespread reduction in active validation of AI outputs (Niederhoffer et al., 2025; Benzing et al., 2025). Theoretically derived claims with corroborating indirect evidence include that these dynamics produce commons-like collective action failures, grounded in Ostrom's empirical work on commons governance and human capital theory's framework for training investment externalities, and that effective Distributed Mastery depends on Internalized Mastery foundations, grounded in distributed cognition theory and empirical evidence on validation capability. Extrapolatory predictions requiring direct empirical test include that expertise depletion will manifest in measurably reduced substantive validation capability across successive cohorts, that depletion effects will be differential across occupations according to the five-factor framework described above, and that governance mechanisms applied at professional-association and policy levels can moderate depletion trajectories. Table 2 proposes indicators, methods, and data sources for each construct as a research agenda through which the framework's predictions can be directly tested.

Four research priorities follow from this mapping. First, longitudinal cohort studies are needed that track expertise development and validation capability across AI adoption periods, following specific professional cohorts from entry to mid-career while documenting the developmental opportunities each cohort accessed or was denied. Second, comparative field studies should examine whether practitioners in AI-heavy versus AI-light organizational contexts show measurable differences in substantive validation capability on domain-specific assessment tasks, drawing on the experimental paradigms developed by Vicente and Matute (2023) and Macnamara et al. (2024). Third, validation failure audits in high-stakes domains would establish whether error rates on AI-generated outputs are increasing over time in ways that would be expected if Internalized Mastery is eroding. Fourth, professional associations should be studied as governance actors, examining whether associations that implement supervised practice requirements or developmental infrastructure standards produce better commons health outcomes than those that do not.

**Table 2**

*Proposed Indicators for Cognitive Commons Health: A Research Agenda*

| Construct | Proposed Indicator | Method | Plausible Data Source |
|---|---|---|---|
| Cognitive Commons Stock | Cohort-specific employment rates at entry levels; proportion of practitioners demonstrating independent task performance across experience cohorts | Labor market analysis; longitudinal cohort study | Payroll and tax records (Brynjolfsson et al., 2025); resume and job posting data (Hampole et al., 2025); professional licensing databases |
| Commons Regeneration Rate | Rate at which entry-cohort practitioners develop independent performance | Difference-in-differences analysis; | Organizational learning records; professional |
| | capability relative to an AI-assisted baseline; entry-level hiring ratios across firms and years | longitudinal cohort tracking | association cohort data; LinkedIn workforce analytics |
| Internalized Mastery | Task performance accuracy under no-AI conditions, stratified by experience cohort and AI exposure duration | Controlled experimental assessment; expert panel structured task evaluation | Occupational certification data; clinical competency assessments; structured expert judgment paradigms |
| Distributed Mastery | AI-assisted task completion accuracy; error detection rate when AI output contains a seeded error; breadth of effective AI application across domain task types | Experimental design with AI-correct vs. AI-incorrect conditions; workflow analysis | Organizational performance records; controlled experimental studies |
| Validation Tether Strength | Rate of detection of plausible-but-incorrect AI outputs, stratified by practitioner experience level and AI exposure duration | Error-detection paradigm (Vicente & Matute, 2023); audit studies comparing senior and junior | Clinical audit data (Budzyń et al., 2025); professional examination records; organizational quality review data |

| | | practitioner detection rates | |
|---|---|---|---|
| Free-Rider Dynamics | Entry-level hiring rate relative to AI adoption rate by firm; firm-level training investment as a function of market share; between-firm variance in developmental infrastructure maintenance | Organizational survey; panel regression with firm and industry fixed effects | SHRM/HRIS survey data; financial reporting (training expenditure); LinkedIn workforce analytics by firm size and sector |

***Note.*** Indicators are proposed as a research agenda, not as validated measurement instruments. Each row identifies the construct, a primary operational approach, an appropriate method, and representative data sources from which commons health could be monitored. The framework predicts that Internalized Mastery and Validation Tether Strength will diverge from Distributed Mastery as AI adoption deepens, making between-construct comparisons a key evidentiary target. Error-detection paradigms following Vicente and Matute (2023) are recommended as the most direct method for validating Validation Tether Strength.

Qualitative research should explore how professionals across career stages experience the regeneration crisis and the tensions between competing forms of expertise development. Are mid-career practitioners finding their Internalized Mastery underutilized as organizations prioritize AI-assisted productivity? Are early-career professionals struggling to access developmental opportunities that build deep domain knowledge? Ardichvili (2022) raised crucial questions about how learning occurs "in the conditions where one's community of learning and practice consists of not only humans but also artificial colleagues" and how to "facilitate social and incidental learning when the pool of learners and co-workers includes artificial agents" (p. 93). Phenomenological studies (Van Manen, 1990) could capture the lived experience of managing contradictory developmental imperatives and how the absence of human mentors and communities of practice affects expertise development trajectories.

From a systems perspective, research should model the feedback loops and time delays that accelerate or decelerate commons depletion, and comparative research across professional domains could identify natural experiments where policy interventions or professional association actions altered depletion trajectories.

**Governance Mechanisms and Collective Action**

The countervailing dynamics reviewed earlier suggest that adaptive responses to AI adoption emerge in some contexts without coordinated governance. Workflow designs that preserve developmental engagement (Everett et al., 2025), worker adaptive capacity (Manning & Aguirre, 2026), and heterogeneity in retrainability across occupations and labor market conditions (Hyman et al., 2025) all indicate that the depletion mechanism operates conditionally rather than universally. The governance arrangements considered below should therefore be read as hypotheses about where and how intervention may be warranted, not as universal prescriptions. These questions also extend to the developmental pathways themselves: whether the entry-level-employment regeneration mechanism analyzed here is the only viable form, or whether alternative arrangements such as revived apprenticeship, AI-mediated mentorship, or distributed credentialing could emerge to perform the same function, is itself an open empirical question. Whether such arrangements are necessary, at what level they should operate, and through what institutional vehicles they might be sustained, are themselves empirical questions the framework generates rather than answers.

The framework points to three levels at which governance arrangements might form: organizational, professional-association, and public policy. Each addresses a different facet of the collective action structure, and each warrants empirical investigation rather than uniform implementation. Organizational practices represent the most immediate level and the most accessible to HRD practitioners. Practices that show promise include phased AI introduction that preserves developmental stages before AI assistance is introduced, cognitive reserve practices that establish minimum periods of independent human performance before AI-augmented work, and AI-restricted learning spaces where early-career professionals engage in deliberate practice without algorithmic assistance. Choudhuri et al.'s (2025) survey of 860 developers found that practitioners resist AI assistance in mentoring and identity-centric tasks, suggesting that workforce preferences may naturally preserve some domains of independent cognitive engagement. Danry et al.'s (2023) experimental research found that framing AI explanations as questions, rather than providing causal explanations directly, improved participants' logical discernment accuracy within session. Designs requiring active reasoning before AI output exposure may, in turn, support independent judgment. Everett et al.'s (2025) trial of physician-AI collaboration found that structuring workflows to require active clinician engagement with AI reasoning, rather than passive acceptance of AI outputs, preserved and even improved diagnostic performance. How organizations design AI integration matters as much as whether they adopt it.

The professional-association level represents the scale at which Ostrom's (1990) governance research is most directly relevant. Her empirical work identified eight design principles characteristic of long-enduring self-organized commons institutions: clearly defined boundaries, proportional equivalence between benefits and costs, collective choice arrangements, monitoring, graduated sanctions (typically operating as reputational or peer consequences rather than state enforcement in self-organized arrangements), conflict resolution mechanisms, recognition of rights to organize, and nesting of local rules within broader systems. Her central finding was that these arrangements emerged through local experimentation and adaptation rather than through external imposition (Ostrom, 1990, 2010). For the Cognitive Commons, this suggests a polycentric approach: professional associations in different fields experimenting with different combinations of these principles, observing outcomes, and propagating arrangements that succeed. Associations in medicine, law, and engineering have institutional infrastructure that could support such experimentation. Those in fields like technology and financial analysis have less developed capacity, and the framework predicts that commons dynamics will manifest most acutely where governance experimentation is least developed.

Public policy represents the third potential level of governance, and the one most relevant to the question Daley (2025) raises from a labor economics perspective. Daley's formal model identifies structural levers that could in principle moderate commons depletion: subsidies to organizations maintaining developmental infrastructure, metrics that reward quality over volume in evaluating professional output, and arrangements that sustain human oversight in AI-intensive work. Whether any of these levers are necessary, or sufficient, in any given occupational context is an empirical question. Where commons vulnerability is high and adaptive responses prove insufficient, possible policy responses emphasize investment rather than prohibition: training investment subsidies in occupations where AI substitution eliminates the productive activity through which expertise traditionally formed; tax credits and matching funds for firms that preserve developmental pipelines; competitive grants for professional associations experimenting with governance arrangements; voluntary tiered credentials that recognize demonstrated Internalized Mastery without restricting AI-assisted practice; and research funding to identify which regeneration mechanisms, traditional or emergent, succeed under AI-integrated work. In domains where safety-critical oversight is already subject to risk-based regulation, existing regulatory instruments can be extended to recognize the Validation Tether, but new regulatory requirements are neither the primary nor the first-order response the framework suggests. These remain hypotheses about institutional design rather than policy recommendations. The framework's contribution is to identify the conditions under which such interventions might be warranted, not to specify which interventions should be adopted.

Governance arrangements must also recognize that Distributed Mastery is a legitimate and valuable form of professional expertise requiring its own developmental infrastructure and standards. The goal of commons governance is not to resist AI adoption or to restore pre-AI developmental pathways as the only legitimate model. It is to ensure that the profession maintains both forms of expertise in functional proportion, with sufficient Internalized Mastery among a cadre of practitioners to provide the substantive validation foundation that effective Distributed Mastery requires. Voluntary tiered credentials might recognize practitioners with demonstrated Internalized Mastery alongside those with AI-assisted competence, enabling market and client differentiation without restricting practice. Continuing education requirements might mandate periodic AI-free performance assessment to ensure that practitioners who predominantly work with AI assistance maintain substantive validation capability. The governance design challenge is to create structures that reward and require investment in Internalized Mastery without stigmatizing the AI collaboration skills that constitute Distributed Mastery.

A systems-level view of training and firm performance (Garavan et al., 2021) converges with McLean's (2004) conception of National HRD as an integrative framework linking individual learning, organizational development, and national competitiveness. The Cognitive Commons framework elevates this convergence from theoretical observation to urgent policy concern. Nations depend on domestic expertise pools for functions that cannot be outsourced during crisis, including critical infrastructure, healthcare, and cybersecurity. The defense industrial base illustrates the stakes: private-sector firms producing national security capabilities recruit from the same civilian professional commons as commercial employers, and commons depletion in fields like software engineering or cybersecurity directly constrains the nation's capacity to maintain technological advantage. If a nation's professional commons erodes through ungoverned market dynamics while a strategic competitor sustains developmental infrastructure through apprenticeship investment, training subsidies, or other incentive structures, the result is a structural capability gap that compounds across cohorts. Comparative international research examining how nations with strong apprenticeship traditions navigate AI adoption differently from those with market-mediated professional development could identify governance models that preserve regeneration capacity without impeding innovation.

**Reconceptualizing HRD Education and Professional Development**

HRD education now prepares professionals for AI-integrated workplaces through generative AI applications in instructional design (Chai et al., 2025), responsible AI training frameworks (Chen, 2024), and technology competence development (Ekuma, 2024). These are necessary competencies, but insufficient if HRD is to steward expertise ecosystems. Current HRD graduate education typically presents the performance paradigm and the learning paradigm as reconcilable through careful design (Swanson & Holton, 2009). The Cognitive Commons reveals that these paradigms sometimes pull in genuinely contradictory directions, and professionals need preparation for that tension.

Beyond AI tool proficiency, HRD professionals need systems thinking about expertise ecosystems and collective action challenges; distributed cognition literacy, including familiarity with how expertise distributes across human-AI networks (Hutchins, 1995; Clark & Chalmers, 1998; Woolley & Gupta, 2024); diagnostic abilities for assessing commons health indicators in their domains; and ethical frameworks for reasoning through situations where organizational performance interests conflict with profession-level commons needs. Most critically, HRD practitioners need strategic influence capabilities to advocate for long-term developmental investment against short-term efficiency pressures, making invisible commons value visible to organizational decision-makers.

## Conclusion: Recognizing the Commons Before the Tragedy

The Tragedy of the Cognitive Commons is not metaphysically inevitable, but it is structurally predictable given current incentive structures and theoretical frameworks. A convergence of empirical evidence indicates that the regeneration mechanism faces disruption in affected sectors: workers ages 22 to 25 in AI-exposed occupations have experienced employment declines while experienced workers remain stable, precisely the cohort pattern that commons depletion through foreclosed regeneration predicts. Ardichvili (2022) warned that automation limits opportunities for developing deep expertise, but the field has lacked a framework for understanding this as a collective action problem requiring profession-level governance rather than organizational-level training solutions.

Without intervention, organizational responses to competitive pressures will deplete the shared expertise that all organizations rely on. By eliminating entry-level positions, organizations prioritize efficiency while socializing the costs of diminishing knowledge. This leads to increased systemic fragility, where failures occur due to validators lacking the Internalized Mastery required for substantive oversight and a loss of meaningful agency in automated work.

As retirees leave without adequate successors, the pool of experts shrinks, reducing the ability for independent validation and crisis management.

HRD can play a central role in addressing this challenge. The field possesses theoretical foundations spanning learning, development, and organizational systems. HRD maintains relationships with organizational decision-makers who shape developmental investments and with professional associations that could coordinate collective action. The field bridges adult education's concern with how humans learn complex capabilities and management's focus on organizational effectiveness.

But embracing commons stewardship requires acknowledging uncomfortable realities that challenge the field's foundational assumptions and professional positioning. Sometimes effective HRD practice cannot harmonize all stakeholder interests. Sometimes serving long-term collective needs means advocating for organizationally suboptimal investments. Sometimes our theoretical foundations (performance versus learning, individual versus collective, efficiency versus resilience) pull in genuinely contradictory directions without resolution. Critical HRD scholars have long recognized these tensions (Bierema & Callahan, 2014; Fenwick, 2004), but the Cognitive Commons makes them unavoidable for all HRD practitioners, not just those adopting critical stances.

This requires HRD to expand its mandate beyond organizational boundaries to encompass profession-level expertise ecosystems. The field must develop governance mechanisms enabling collective action despite free-rider incentives, prepare practitioners to contend with genuine tensions rather than assume false harmony, and advocate for developmental pathways that maintain both Distributed Mastery and Internalized Mastery. These imperatives will shape not just the field of HRD, but the future relationship between human and artificial intelligence in organizations and society.

The Tragedy of the Commons need not be inevitable. Ostrom (1990) demonstrated that with appropriate governance, commons can be sustained, often through arrangements that emerge from local experimentation rather than central design. Sustaining a commons first requires recognizing that one exists, understanding the collective action challenges threatening it, and coordinating responses that overcome free-rider incentives. For the Cognitive Commons, HRD can catalyze recognition, contribute to understanding, and inform coordination. The field emerged to develop human capability in service of organizational effectiveness. It can now expand its scope to consider collective expertise as a resource that warrants stewardship. The question is whether HRD will engage this challenge or remain confined by assumptions that no longer adequately describe the world we inhabit.